\documentclass[aps,prb,reprint,groupedaddress
]{revtex4-1}
\usepackage{bm,amssymb,graphicx}
\def\be{\begin{equation}}  
\def\ee{\end{equation}}  
\def\ba{\begin{eqnarray}}  
\def\ea{\end{eqnarray}}  
\def\bc{\begin{center}}  
\def\ec{\end{center}}  
  
\begin{document}


\title{Reply to the Comment on ``Theory of microwave-induced zero-resistance states in two-dimensional
electron systems'' and on ``Microwave-induced zero-resistance states and
second-harmonic generation in an ultraclean two-dimensional electron gas'' }


\author{S.~A.~Mikhailov}
\affiliation{Institute of Physics, University of Augsburg, D-86161 Augsburg, Germany}



\date{\today}

\begin{abstract}
We show that all questions raised in the Comment can be easily answered within the theory formulated in the commented papers. It is also shown that the bulk theory promoted by the commenter fails to explain the most important fundamental features of the discussed phenomenon.
\end{abstract}


\maketitle


In the Comment \cite{ZudovComment} Zudov claims that our theory \cite{Mikhailov11a,Mikhailov14a} does not explain the microwave induced resistance oscillations/zero resistance states (MIRO/ZRS) and that MIRO/ZRS is a bulk phenomenon. In this Reply we show that, focusing his attention on only a few very specific aspects of the effect, Zudov does not notice a number of global, fundamental problems of MIRO/ZRS, which could be understood only after assuming that this phenomenon has a near-contact origin \cite{Mikhailov11a,Mikhailov14a}. 

The MIZRS phenomenon \cite{Mani02,Zudov03} has demonstrated \textit{at least five} mysterious features which seemed to contradict both the previously known physics and common sense:

\textbf{1.} The MIZRS effect is huge. The photoresistance maxima were found to be $7-10$ times larger than the dark resistance values. The photoresistance is a nonlinear phenomenon: the dc current is proportional to the dc electric field $E_0$ and the squared microwave field $E_\omega$, $j\propto E_{dc}E_{\omega}^2$. As known,  substantial nonlinear effects can be observed only if the corresponding electric field parameter exceeds unity: 
\be 
{\cal F}\equiv \frac{eE_\omega}{\omega p_F}=\frac{eE_\omega v_F}{2\omega E_F}=\frac{eE_\omega }{\hbar\omega \sqrt{2\pi n_s}}\gtrsim 1. \label{conditionField}
\ee
Physically ${\cal F}$ is the additional momentum $\sim eE_\omega/\omega$ (energy $\sim eE_\omega v_F/\omega$), acquired by an electron during one oscillation period, normalized to its average momentum $p_F$ (energy $E_F$); here $\omega$ is the microwave frequency, $p_F$, $v_F$, $E_F$ are the Fermi momentum, velocity and energy, and $n_s$ is the density of two-dimensional (2D) electrons. The condition (\ref{conditionField}) can be rewritten in terms of the microwave power density ${\cal P}$ required for the observation of such a huge MIZRS effect:
\be 
{\cal P}\gtrsim {\cal P}_0 = \left(\frac{\hbar\omega }{e}\right)^2 n_s c.
\label{conditionPower}
\ee
For typical MIZRS parameters ($f\simeq 100$ GHz, $n_s\simeq 3\times 10^{11}$ cm$^{-2}$) the required nonlinear power density (\ref{conditionPower}) is 
\be 
{\cal P}_0\simeq 1.7\textrm{ kW/cm}^2.
\ee
In real MIZRS experiments the microwave power density was \textit{about six orders of magnitude smaller} ($\lesssim 1$ mW/cm$^2$, Refs. \cite{Mani02,Zudov03}). The first MIZRS puzzle was thus:  
\textit{How such a huge effect can be observed at so low microwave powers?}

\textbf{2.} The MIRO/ZRS effect demonstrates strong oscillations not only around the fundamental cyclotron frequency $\omega=\omega_c$ but also around harmonics $\omega=n\omega_c$, $n=2,3,4,5,\dots$. It is well known that in the uniform external ac electric field the transitions between Landau levels $E_N=\hbar\omega_c(N+1/2)$ are forbidden if $\Delta N\neq \pm 1$. In order to violate this selection rule the microwave field should be strongly inhomogeneous on the cyclotron radius ($r_c$) scale: the nonlocal bulk conductivity of the 2D electron gas in the magnetic field $B$ has the form \cite{Chiu74}
\be 
\sigma_{xx}(\omega,q)= \frac{n_se^2i\omega}{m^\star}\sum_{n=1}^\infty \frac{[2nJ_n(X)/X]^2}{\omega^2-(n\omega_c)^2}
\label{sigmaXX}
\ee
where $X=qv_F/\omega_c=qr_c$ is the non-locality parameter, $q$ is the wave-vector of the external wave and $J_n$ are Bessel functions. In the MIRO/ZRS experiments the radiation wavelength ($\simeq 3$ mm) and the sample dimensions ($\simeq 0.2-1$ mm) were much larger than the cyclotron radius ($r_c\simeq 1-5$ $\mu$m). As seen from (\ref{sigmaXX}), the amplitude of the $n$-th resonance (at $\omega=n\omega_c$) scales at $qr_c\ll 1$ as $A_n\propto[n(qr_c/2)^{n-1}]^{2}$, which for realistic experimental parameters ($qr_c\lesssim 10^{-2}$) gives $A_1:A_2:A_3\simeq 1:10^{-4}:10^{-8}$. The experiment shows, however, that the amplitudes $A_n$ fall down very slowly, roughly, like $A_n\simeq 1/n$. Thus the second MIZRS puzzle was: 
\textit{Why the MIZRS effect is so strong at the cyclotron harmonics with $n=2,3,\dots$?} 

\begin{figure}[h!]
\includegraphics[width=\columnwidth]{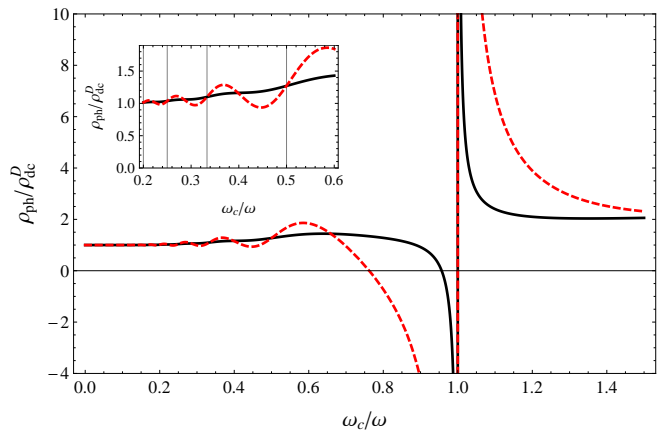}
\caption{Results of the bulk theory \cite{Dmitriev05} of MIRO/ZRS (the ``displacement'' model) plotted for \textit{experimentally realistic} values of the microwave power (black curve) and for the power \textit{ten times larger than in  experiments} (red dashed curve); in the original publication \cite{Dmitriev05} (Fig. 2 there) the microwave power was assumed to be $100-500$ times larger than in reality. Inset shows the enlarged picture around harmonics $\omega/\omega_c=2,3,4,5$. 
\label{fig:my}}
\end{figure}

The bulk theory \cite{Dmitriev05}, which Zudov considers to be the only correct theory of MIRO/ZRS, is not able to overcome these difficulties. Figure \ref{fig:my} shows the $B$-dependence of the microwave induced photoresistance calculated according to Eq. (15) (the ``central result'') of Ref. \cite{Dmitriev05}. For the dimensionless power parameter, which enters this formula, we have chosen the value 0.005 which corresponds, according to a discussion in \cite{Dmitriev05} (Section IX there), to realistic experimental values at the frequency $f=100$ GHz (the power density $<1$ mW/cm$^2$, Refs. \cite{Mani02,Zudov03}). One sees that, in full compliance with the above estimates, the theory \cite{Dmitriev05} does not reproduce \textit{any} oscillations around harmonics $\omega=n\omega_c$, $n>1$. Even if to increase the power \textit{by a factor of ten} (red curves in Fig. \ref{fig:my}) the effect is extremely small. Sufficiently large photoresistance oscillations were obtained in \cite{Dmitriev05} (Fig. 2 there) only when the microwave power was ``theoretically'' increased by a factor of $100-500$ as compared to the experimental values. The still remaining very large difference between the 1st and higher cyclotron harmonics was hidden in \cite{Dmitriev05} by showing oscillations only at $\omega_c/\omega<0.7$. 

\textbf{3.} The third puzzle of MIRO/ZRS is: \textit{Why the effect is insensitive to the circular polarization sense?} All bulk scenarios of the phenomenon predicted a strong difference of the photoresistance response to the clockwise and counterclockwise circular polarization of radiation. The experiment \cite{Smet05} demonstrated that, while the absorption (the bulk effect) is substantially different for different circular polarizations, the MIRO/ZRS effect is practically independent of this. 

\textbf{4.} The MIZRS effect was discovered \cite{Mani02,Zudov03} in samples with an extremely high electron mobility $\mu \sim (15-20)\times 10^6$ cm$^2$/Vs. The same effect measured in 1993 in samples with $\mu \sim 1\times 10^6$ cm$^2$/Vs, Ref. \cite{Vasiliadou93}, showed no MIRO/ZRS (all other parameters in Refs. \cite{Vasiliadou93} and \cite{Mani02} were the same). The fourth MIZRS puzzle is thus: \textit{Why the MIRO/ZRS effect was mostly observed in extremely clean samples?} 

\textbf{5.} It is known that the external electric field $E_{\mathrm{external}}$ is screened in the finite-size samples due to the presence of sample edges. The field $E_{\mathrm{in\ sample}}$ really acting on electrons of the 2D gas is 
\be 
E_{\mathrm{in\ sample}}\simeq E_{\mathrm{external}}\frac{\omega^2-\omega_c^2}{\omega^2-\omega_c^2-\omega_p^2}, 
\label{screened}
\ee 
where $\omega_p\propto 1/\sqrt{w}$ is the 2D plasmon frequency and $w$ is the sample width, see, e.g. Ref. \cite{Mikhailov04a}. One of the consequences of this fact is that the CR is never seen at exactly the cyclotron frequency $\omega_c$ but is shifted to the magnetoplasmon frequency $\omega_{mp}=\sqrt{\omega_c^2+\omega_p^2}$. This so called depolarization shift was many times observed not only in the absorption spectra, but also in the already mentioned microwave photoresistance experiment \cite{Vasiliadou93}. Under the conditions of the MIRO/ZRS experiments the term $\omega_p^2$ \textit{was never small} as compared to $\omega_c^2$ (sometimes it was even larger). Therefore, trying to explain the MIZRS effect by a bulk mechanism one cannot simply ignore the fact that the samples have finite dimensions. But, if to take this fact into account, the external electric field had to be replaced by the screened field (\ref{screened}) in all bulk-theory formulas. This would lead to a very strong shift of all MIRO/ZRS oscillations from the cyclotron to magnetoplasmon resonances. This is not however the case in the experiments. Therefore, the fifth puzzle of the MIZRS phenomenon is: 
\textit{Why the depolarization shift is not seen in the MIZRS oscillations?}

The MIRO/ZRS phenomenon could be completely understood if and only if reasonable answers were found to \textit{all the above raised questions}. No bulk theory could cope with this problem so far.

In the papers \cite{Mikhailov11a,Mikhailov14a} we have shown that \textit{all five above listed misterious features of MIRO/ZRS can be easily explained} if to assume that the origin of the MIRO/ZRS effect lies near the contacts to the 2D gas: 

\textbf{1.} Near the contacts (or near other sharp metallic objects close to the 2D electron gas) the local ac electric field can be substantially larger than the incident-wave field due to the screening of radiation in the metal (the ``lightning rod'' effect). As a result, the field parameter (\ref{conditionField}) turns out to be much larger near the contact than in the bulk of the 2D gas. 

\begin{figure}[h!]
\includegraphics[width=\columnwidth]{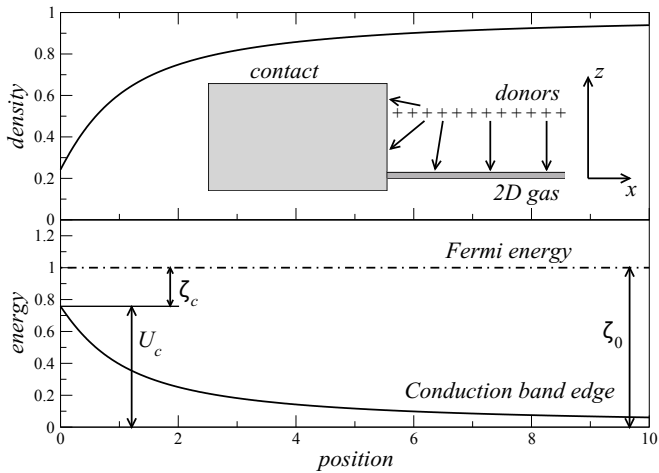}
\caption{A presumptive behavior of the electron density and the conduction-band edge at the boundary 2D electron gas -- contact (in the absence of the magnetic field and microwaves). The inset illustrates a possible mechanism of the Schottky-type barrier formation: if the Fermi energy in the metallic contact is lower than the energy levels of donors, electrons from donors will move (near the contact) to the 3D metal, rather than to the 2D gas (shown by arrows in the inset). As a result, the density of electrons near the contact is reduced and their potential energy is increased by the contact potential difference $U_c$. The near-contact chemical potential $\zeta_c$ is then smaller (or much smaller) than in the bulk $\zeta_0$. \label{fig:bandedge}}
\end{figure}

In addition (this point was not discussed in Refs. \cite{Mikhailov11a,Mikhailov14a}), the contact potential difference between the 2D gas and the contact does not need to be exactly zero (in the absence of the magnetic field and microwaves), therefore, it is quite possible that near the contact the conduction-band edge is bent up forming a Schottky-type barrier, Figure \ref{fig:bandedge}. The bulk electron density, the Fermi momentum and energy in Eqs. (\ref{conditionField}) -- (\ref{conditionPower}) should then be replaced by the corresponding (smaller) near-contact values. This resolves the first MIRO/ZRS puzzle.  

\textbf{2.} Near the contacts the electric field is strongly inhomogeneous on the scale  $\sim 100$ nm (the metal thickness, the distance between the 2D gas and the doping layer, etc.). This is shorter than the cyclotron radius, which explains MIRO at higher CR harmonics. 

\textbf{3.} Near the contacts the electric field of the microwave is linearly polarized perpendicular to the metallic edge  \cite{Mikhailov06a} independent of the circular polarization sense of the incident wave. 

\textbf{4.} The ponderomotive potential $U_{pm}$ oscillates as a function of $\omega/\omega_c$. In order to observe the MIZRS the amplitude of these oscillations should exceed the value $\zeta_c$, see Fig. \ref{fig:bandedge}. In samples with a higher electron mobility the amplitude of $U_{pm}$ oscillations is bigger, therefore the effect is more easily seen in ultraclean samples.  

\textbf{5.} The depolarization shift problem can also be easily resolved by assuming the near-contact origin of MIRO/ZRS. The experimentally measured photoresponse is actually a superposition of two different contributions: bulk and near-contact. The magnetoplasmon resonance is caused by the screening of the external electric field by bulk 2D electrons. The MIRO/ZRS effect results from the screening of the external field by electrons in contacts. Since the density of electrons in metals is orders of magnitude larger than in the 2D gas, the MIRO/ZRS effect is much stronger (in high-mobility samples), and the weak magnetoplasmon resonance turns out to be hidden behind the very strong MIRO. In low-mobility samples MIRO are absent, and one easily sees magnetoplasmons \cite{Vasiliadou93}. This interpretation agrees with some experiments (e.g. Ref. \cite{Yang06}) in which the MIRO/ZRS effect was controllably suppressed and one could see the ``rising'' magnetoplasmon resonance.

The idea of the near-contact origin of MIRO/ZRS phenomenon thus resolves all the misteries of the effect outlined above. Its development in Refs. \cite{Mikhailov11a,Mikhailov14a} showed that specific dependencies of the measured photo-resistance on the magnetic field are caused by ponderomotive forces which attract/repel electrons to/from the contacts, dependent on the ratio $\omega/\omega_c$. 

In this Reply we introduce two further natural modifications of the theory \cite{Mikhailov11a,Mikhailov14a} which allows us to get even better description of experimental data. First, we will assume that the conduction-band edge is bent up near the contacts and, hence, the local chemical potential $\zeta(x)$ is position dependent, with $\zeta_c<\zeta_0$, Figure \ref{fig:bandedge}. Second, in accordance with experimental data (e.g. Ref. \cite{Hatke09}) the scattering rate $\gamma=1/\tau$ will be assumed to be temperature dependent, $\gamma=\gamma(T)$ (in our previous papers Refs. \cite{Mikhailov11a,Mikhailov14a} we considered it to be constant for simplicity).

Now consider  specific criticisms of the Comment.

\begin{figure}[h!]
\includegraphics[width=\columnwidth]{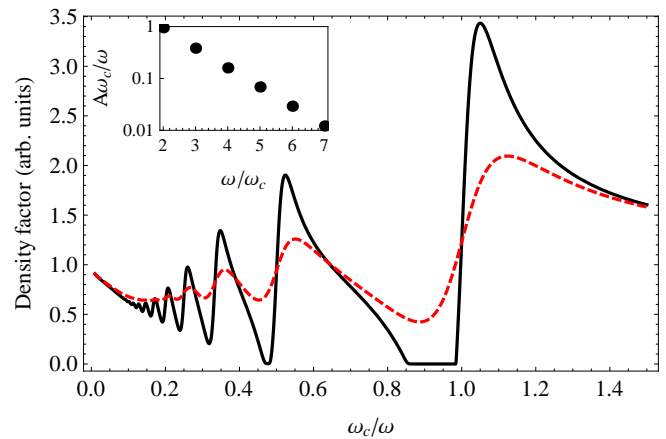}
\caption{$B$ dependencies of MIRO, calculated according to the ponderomotive-force theory \cite{Mikhailov11a,Mikhailov14a}, at the scattering parameters $\gamma(T)/\omega$ differing by the factor of 2.5 (due to the temperature growth from 1 to 5.5 K, according to the experimental data of Ref. \cite{Hatke09}; the red curve corresponds a larger $\gamma$). All parameters are the same as in Figure 8(a) in Ref. \cite{Mikhailov14a}. Inset shows the $\omega/\omega_c$-dependence of the value $A\omega_c/\omega$, where $A$ is the amplitude of calculated \cite{Mikhailov11a,Mikhailov14a} MIRO oscillations (for the black curve). \label{fig:temp}}
\end{figure}

\textit{Temperature dependence.} The reduction of MIRO oscillations with temperature \cite{Hatke09} can be easily explained if to assume that the scattering rate $\gamma(T)$ depends on $T$. In Ref. \cite{Hatke09} the authors observed, apart from the decrease of MIRO amplitudes, also a monotonic growth of the zero-field resistivity (by a factor of $\approx 2.5$ at the temperature change from 1 K to 5.5 K) which they interpret in terms of the temperature induced reduction of mobility due to excitation of acoustic phonons \cite{Stormer90}. Figure \ref{fig:temp} shows the corresponding change of MIRO calculated within the theory \cite{Mikhailov11a,Mikhailov14a}. The results perfectly agree with experiments.

\textit{Sublinear power dependence of MIRO maxima.} The same heating effect ($\gamma(T)$) easily explains the sublinear power dependence of MIRO maxima which the commenter critisizes with a reference to his paper \cite{Hatke11c} (in the reference [74] of the paper \cite{Hatke11c} the authors directly indicate that the zero-field resistivity grew due to the heating at large microwave powers). 

\textit{``Activation'' behavior of  MIRO minima.} According to Refs. \cite{Mikhailov11a,Mikhailov14a} the ``activation'' behavior of the photoresistance is observed when the ponderomotive potential is positive, $U_{pm}\propto {\cal P}>0$, and its amplitude exceeds $\zeta_0$. Then the density factor assumes the form
\be 
{\cal N}=\frac T{\zeta_0}\ln\left[1+\exp\frac{\zeta_0-U_{pm}}T\right]\approx 
\frac T{\zeta_0}\exp\frac{\zeta_0-U_{pm}}T,\label{activat}
\ee
and the dependence $\propto e^{-a{\cal P}/T}$ is evident; here $a$ is a prefactor which depends on $B$, $\omega$, etc.\cite{Mikhailov11a,Mikhailov14a} (we believe that it is misleading to present the ``activation'' behavior of $R_{xx}$ in the form $\exp(-\Delta/T)$, like it is done in the Comment, since this form traditionally implies something like ``opening of a gap $\Delta$'' due to some fundamental interactions; in the MIRO/ZRS phenomenon the dependence $e^{-a{\cal P}/T}$ simply arises because the ponderomotive potential builds a barrier at the boundary 2D gas -- contact, and the electron gas becomes non-degenerate in the near-contact region). The factor $T/\zeta_0$ is very small which leads indeed to some quantitative disagreements with experiments; here $\zeta_0$ is the chemical potential in the bulk of the sample, Figure \ref{fig:bandedge}. Taking into account the quite natural assumption on the band bending in the near-contact region, Figure \ref{fig:bandedge}, we see that the parameter $T/\zeta_0$ should be replaced by $T/\zeta_c$ where the near-contact value of the chemical potential  $\zeta_c=\zeta_0-U_c$ is smaller than in the bulk by the contact potential difference $U_c$. This removes the contradictions mentioned in the Comment. 

\textit{The ``phase'' of MIRO.} A somewhat larger value of $\gamma/\omega$ which was needed for the interpretation of ``phases'' (for the definition of ``phases'' see the Comment) can be explained by the fact that the CR linewidth in the very high mobility samples contains, in addition to $\gamma$, the (much larger) radiative decay contribution, see a detailed discussion in Ref.\cite{Mikhailov04a}. It is worth noting that the bulk theory \cite{Dmitriev05} demonstrates an extremely small ``phase'' $\delta_1$, see Figure \ref{fig:my}, and tiny oscillations around $\omega\simeq n\omega_c$. However, this  does not impede the commenter to insist on the validity of this theory.

\textit{Dependence on magnetic field.} In order to check whether the theory \cite{Mikhailov11a,Mikhailov14a} reproduces the experimentally observed $B^{-1}\exp(-B_0/B)$ dependence of the MIRO amplitudes we have measured the MIRO amplitudes $A$ in Figure \ref{fig:temp} (the black curve) and plotted $A\omega_c/\omega$ versus $\omega/\omega_c$ in the Inset to Figure \ref{fig:temp}. One sees that the theory \cite{Mikhailov11a,Mikhailov14a} perfectly agrees with experimental data. 

\textit{Dependence on in-plane magnetic field.} As seen from Figure \ref{fig:my}, the formulas which the commenter uses for interpretation of his experiments, underestimate the effect by several orders of magnitude, therefore his attempt to explain the $B_\parallel$-dependence of MIRO/ZRS by the corresponding dependence of a ``quantum lifetime'' (one of numerous fitting parameters of the bulk theory) does not seem to be convincing. 

Since all other (so far mysterious) MIRO/ZRS features have been successfully explained by the ponderomotive-force theory \cite{Mikhailov11a,Mikhailov14a}, we believe that the origin of the $B_\parallel$ dependence should be also searched for in the near-contact regions. For this one needs more information about geometry and physical properties (e.g. the work function) of contacts. In available experimental publications on MIRO/ZRS this information is, however, missing. 

\textit{Dependence on dc electric field.} That MIRO/ZRS is \textit{not} a bulk phenomenon is evident from the above discussion. As for the influence of the dc electric field on MIRO/ZRS, this problem has not been treated in the commented papers \cite{Mikhailov11a,Mikhailov14a} at all, therefore any discussion of these effects both in the Comment and in this Reply is irrelevant. 

\textit{Conclusions:}

\textbf{1.} The bulk theories fail to explain five main misterious features of MIRO/ZRS listed in the beginning of this work. The seeming agreement with experiments was achieved by using in calculations $100-500$ times larger microwave power values, as compared to experimental ones.  

\textbf{2.} The ponderomotive-force theory \cite{Mikhailov11a,Mikhailov14a}, which assumes the near-contact origin of the effect, naturally explains all MIRO/ZRS puzzles and answers the questions raised in the Comment. In contrast to bulk theories which introduced a large number of different scattering times (fitting parameters) the theory \cite{Mikhailov11a,Mikhailov14a} consistently describes all experimentally observed features by using only one scattering parameter -- the momentum relaxation rate $\gamma$. 

I thank Nadja Savostianova for reading the manuscript and useful comments.

%

\end{document}